\documentclass[tradiabstract]{aa} 

\usepackage{graphicx,txfonts}
\usepackage{amsmath}
\usepackage{mathtools}
\usepackage{relsize}

\newcommand{\Lsun} {L$_{\odot}$}
\newcommand{\Msun} {M$_{\odot}$}

\newcommand{\simgreat} {\mathbin{\lower 3pt\hbox{$\rlap{\raise
        5pt\hbox{$\char'076$}}\mathchar"7218$}}}

\newcommand{\simless}{\mathbin{\lower 3pt\hbox {$\rlap{\raise
        5pt\hbox{$\char'074$}}\mathchar"7218$}}}

\begin{document}

\title{Far-infrared emission of massive stars}

\authorrunning{Siebenmorgen, Scicluna, Kre{\l}owski}
\titlerunning{Far infrared emission of massive stars}

\author { R.~Siebenmorgen\inst{1}, P.~Scicluna\inst{2}, J. Kre{\l}owski\inst{3}}

\institute{European Southern Observatory, Karl-Schwarzschild-Str. 2, 85748
  Garching b. M\"unchen, Germany, Ralf.Siebenmorgen@eso.org 
\and
Institue of Astronomy and Astrophyscis, Academia Sinica, 11F of AS/NTU
Astronomy-Mathematics Building, No.1, Sec.~4, Roosevelt Rd, Taipei
10617, Taiwan, R.O.C.
\and
Center for Astronomy, Nicolaus Copernicus University, Grudzi{\c{a}}dzka
5, Pl-87-100 Toru{\'n}, and Rzesz{\'o}w University, al. T. Rejtana 16c, 35-959 Rzesz{\'o}w, Poland
}

\date{Received 01/06/2018, Accepted 04/09/2018}

\abstract{We present { results of the analysis of} a sample of 22
  stars of spectral types from O7 to B5 and luminosity classes I--V
  for which  spectra from the Infrared Spectrograph (IRS) of Spitzer
  are available. The IRS spectra of these stars are examined for signs
  of excess infrared (IR) emission by comparison with{ stellar
    atmospheric spectra}.  We find that the spectra of half of the
  studied stars are dominated by excess emission in the far-IR,
  including all six super- and bright giants. In order to examine the
  origin of the far-IR excess, we supplement the Spitzer data with
  optical high-resolution echelle spectroscopy ($\lambda/\Delta
  \lambda \sim 10^5$), near-IR high-contrast coronagraphic imaging
  taken with the SPHERE instrument at VLT with a spatial resolution of
  $0\farcs$05, and WISE and Herschel photometry.  In the optical
  region, we detect various absorption and emission lines (H~$\alpha$,
  C~III, and N~III) irrespective of the far-IR excess. Pfund~$\alpha$
  and Humphrey~$\alpha$ lines are observed at the same time as the
  far-IR { excess. These lines are stronger in stars with far-IR
    excess than in stars without excess.} A scattered-light disk in
  the central $r \simless 2.5 \arcsec$ region of{ the far-IR excess
    stars } HD~149404, HD~151804, and HD~154368 can be excluded from H
  band imaging down to a 1$\sigma$ contrast of $F(r)/F_{*} \sim
  10^{-6}$.  The far-IR excess is fit either by a free-free component
  from ionized gas as for the winds of hot stars or a{ large (1\,pc)}
  circumstellar dust shell. The putative dust envelopes required to
  explain the excess have a visual extinction as low as a few hundred
  $\mu$-mag.}

\maketitle

\section{Introduction}
Thanks to the sensitivity of recent space missions, particularly the
Spitzer and Herschel space telescope, infrared (IR) excess emission has been
revealed for a large number of stars in a variety of environments.  In
low-mass stars, IR excess emission is commonly attributed to the
presence of circumstellar dust, which absorbs stellar radiation and
re-radiates it in the IR.  This includes protoplanetary disks and AGB
shells as well as debris disks.  In more massive main sequence stars,
excess emission is commonly attributed to thermal free-free emission
in a hot, dense ionised wind (Hartmann \& Cassinelli 1977). This
produces a continuum of emission from the IR to the radio with a
distinctive power-law spectral index $f{\nu} \propto \nu^{0.7}$
(Barlow 1979). A particular case are Be-stars emitting copious
free-free radiation (Rivinius al. 2013 for a recent review). However,
in the IR it is difficult to distinguish free-free from optically thin
dust emission.  The presence or absence of dust could hold important
information regarding sources of interstellar dust, particularly in
high-redshift galaxies, and the formation and evolution of massive
stars.

We present a sample of massive stars for which Spizter/IRS (Houck et
al. 2004) 5 - 35\, $\mu$m spectra are available.  These mid- to far-IR
spectra{\footnote {We use terminology of near-IR including atmospheric
    windows in the JHK bands, mid-IR including LMN bands, and far-IR
    for longer wavelengths up to 200$\mu$m}} are used to search for
excess emission over the expected photospheric flux, which we seek to
correctly attribute to wind and/or dust emission. In Sect.~2 we
outline our sample selection and the data available for the
stars. Section~3 discusses the modeling of the excesses as either wind
or dust emission including the possible origins of the
grains. Section~4 summarises our conclusions.

\begin{table*} [h!tb]
\begin{center}
  \caption{Details of sample stars. We specify coordinates, spectral
    type, visual magnitude and extinction along the sightline, and
    reddening. { Stars where we detect far-IR excess emission are
      marked in bold.}
    \label{sample.tab}}
\begin{tabular}{llllcc}    
  \hline
    \hline
& & & &  &  \\    
Designation      & RA & DEC & {SpT} & V & $E(B-V)$ \\
  &\multicolumn{2}{c}{J2000} & & (mag) & (mag) \\
  \hline
{\bf HD~24912}  & 03:58:57.90 &$+$35$^{\rm o}$ 47$'$27.71$''$   & O8IV   &4.06  &0.35 \\
HD~34816  &05:19:34.52 & $-$13$^{\rm o}$10$'$36.44$''$  & B0.5IV &4.29  &0.05 \\
HD~36861  &05:35:08.27 & $+$09$^{\rm o}$56$'$02.96$''$  & O8III  &3.47  &0.09 \\
HD~38087  &05:43:00.57& $-$02$^{\rm o}$18$'$45.38$''$   & B3II   &8.29  &0.31 \\
HD~47839  &06:40:58.66& $+$09$^{\rm o}$53$'$44.72$''$    & O7V    &4.64  &0.07 \\
HD~53367  &07:04:25.53 &$-$10$^{\rm o}$27$'$15.74$''$   & B0IV{ e}&6.96    &0.74 \\
HD~62542  &07:42:37.22 &$-$42$^{\rm o}$13$'$47.84$''$   & B3V &8.03      &0.36 \\
{\bf HD~64760}  &07:53:18.16 &$-$48$^{\rm o}$06$'$10.56$''$     & B0.5III &4.24 &0.08 \\
HD~65575  &07:56:46.71 &$-$52$^{\rm o}$58$'$56.47$''$   & B3IV   &3.43  &0.05 \\
HD~108639 &12:29:09.51 &$-$60$^{\rm o}$48$'$17.55$''$   & B1III  &8.57  &0.35 \\
HD~147933 &16:25:35.10 &$-$23$^{\rm o}$26$'$48.70$''$   & B1.5V  &5.05  &0.47 \\
{\bf HD~149404} &16:36:22.56 &$-$42$^{\rm o}$51$'$31.90$''$     & O8.5Ia &5.52  &0.62 \\
{\bf HD~149757} &16:37:09.54 &$-$10$^{\rm o}$34$'$01.53$''$    & O9.2V &2.56   &0.31 \\
{\bf HD~151804} &16:51:33.72 &$-$41$^{\rm o}$13$'$49.93$''$     & O8Iab&5.22    &0.30 \\
{\bf HD~154368} &17:06:28.37 &$-$35$^{\rm o}$27$'$03.77$''$     & O9.5Ia &6.13  &0.76 \\
{\bf HD~165024} &18:06:37.87 &$-$50$^{\rm o}$05$'$29.31$''$     & B4II &3.66    &0.05  \\
{\bf HD~188209} &19:51:59.07 &$+$47$^{\rm o}$01$'$38.42$''$     & O9.5III &5.63 & 0.15 \\ 
{\bf HD~190918} &20:05:57.32 &$+$35$^{\rm o}$47$'$18.15$''$     & O9.7Ia &6.75  & 0.41 \\
{\bf HD~192639} &20:14:30.43 &$+$37$^{\rm o}$21$'$13.81$''$     & O7.5Iab &7.11 & 0.61  \\
{\bf HD~206773} &21:42:24.18 &$+$57$^{\rm o}$44$'$09.79$''$    & B0V{ e} &6.87     &0.45  \\
{\bf HD~210839} &22:11:30.58 &$+$59$^{\rm o}$24$'$52.15$''$     & O6Ib &5.05    & 0.57 \\
HD~214680 &22:39:15.68 &$+$39$^{\rm o}$03$'$00.97$''$   & O9.5V  &4.88  &0.08 \\
  \hline
\end{tabular}
\end{center}

{ Notes. HD~38087 is a binary, HD~53367 is a Herbig Be binary,
  HD~190918 is a WR+O star binary, HD 206773 is a Be star, HD~149404
  is a multiple object, HD~214680 shows no sign
  of a binary in our two HR spectra despite that it is marked as a binary in SIMBAD.}

\end{table*}

\section{Observations}

\subsection{Sample}
  
We select a sample of targets that forms a subsample of the
Voshchinnikov \& Henning (2010) sample of 196 OB stars, which were
originally selected for knowledge of the dust phase abundances of O,
Mg, Fe and Si in the interstellar medium (ISM). The available data was
extended in Voshchinnikov et al. (2012) by collating fractional
polarisation measurements. In the optical region it is shown that the
wavelength dependence of polarisation towards these stars follows a
Serkowski curve (Bagnulo et al. 2017). The polarisation cannot
therefore be due to scattering on dust or free electrons and { must
  come from} dichroic absorption { by} aligned dust
particles. Further, it is assumed that the dichroism appears along the
sightline towards the star in the diffuse ISM (Siebenmorgen et
al. 2018), and ignoring any connection with the circumstellar
environment of the stars. Voshchinnikov et al. (2012) study the role
of the elements in the grain alignment that results in polarisation.
Table~\ref{sample.tab} lists the relevant details of our sample, taken
from the aforementioned papers, and references to the spectral types
of the star.

\subsection{Archival data}

Our selection of stars is based on the availability of archival IRS
(Houck et al. 2004) spectra, giving us a total of 22 stars. These data
were complimented with optical (UBVRI) and near-IR (JHK) photometry
obtained from { Vizier/CDS} and the 2MASS point-source catalogue
(Cutri et al. 2003; Skrutskie et al. 2006). Where possible, further
archival IR data were collected including those from the Infrared
  Array Camera (Fazio et al. 2004) of Spitzer, AKARI (Ishihara et al.
  2010), the Infrared Astronomical Satellite (IRAS, Neugebauer et
  al. 1984; Joint Iras Science 1994), the Wide-Field Infrared Survey
  Explorer (WISE, Cutri et al. 2012) photometry, and Infrared Space
  Observatory (ISO) spectra from the Short Wavelength Spectrometer
  (SWS, Sloan et al. 2003). In most cases only upper limits are
available. Unfortunately it has not been possible to locate
higher-quality data at wavelengths longer than 40\,$\mu$m from the
Multi-Band Imaging Photometer (MIPS) of Spitzer. However, point-source
photometry of the Photodetector Array Camera and Spectrometer (PACS)
of Herschel (Marton et al., 2017) is available at 70, 100, and
160\,$\mu$m for { HD~24912, HD~149404, HD~149757, HD~151804, and
  HD~210839}.

\begin{figure*} [h!tb]
\includegraphics[width=18.35cm,clip=true,trim=0cm 0.cm 1cm 3.cm]{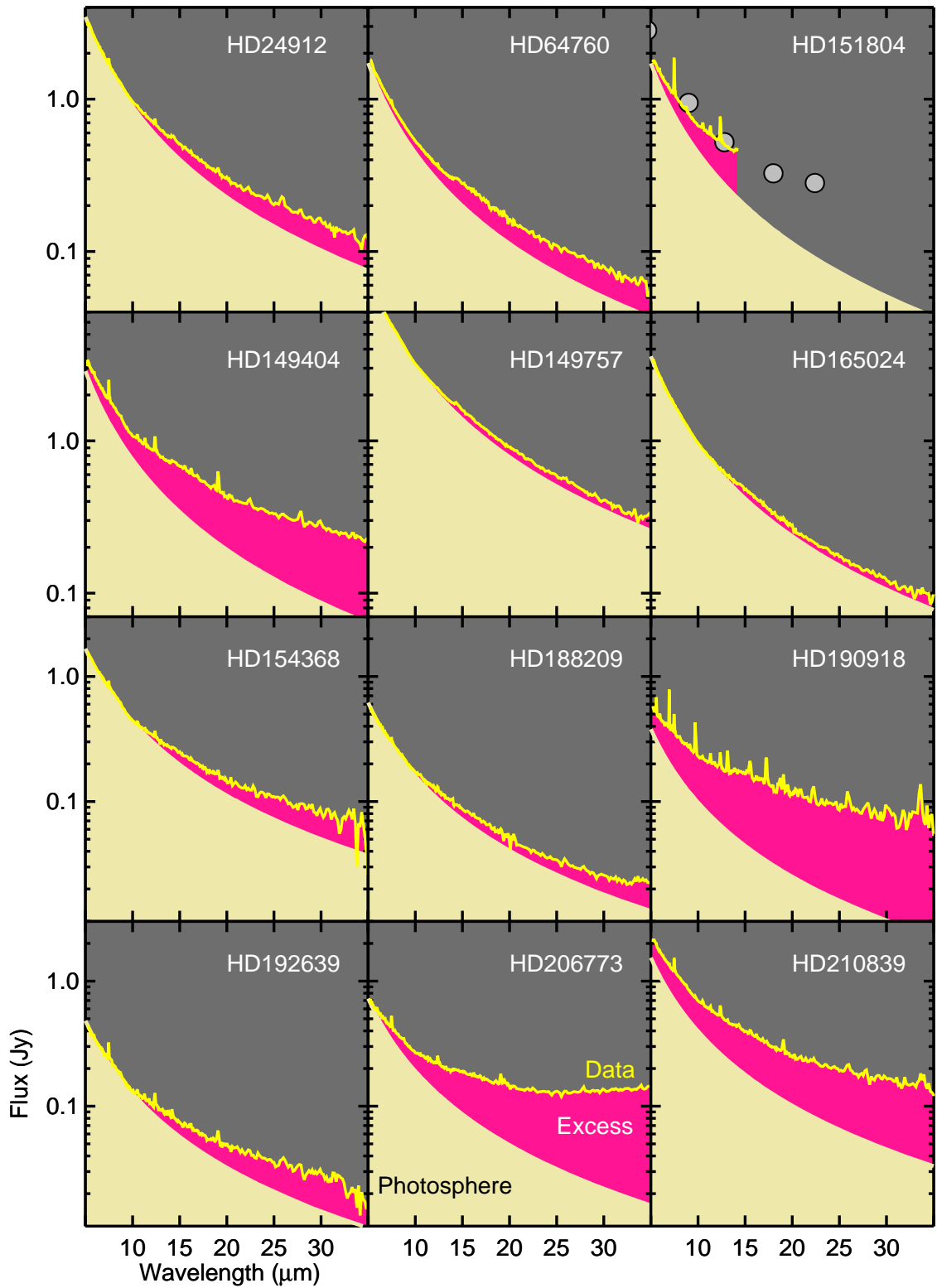}
\caption{Infrared (5--35\,$\mu$m) SEDs of sources. The yellow shaded
  region shows the photospheric contribution, while the yellow line
  shows the IRS spectrum. The magenta shaded region in between
  highlights the excess. Photometry for HD~151804 are shown as grey
  circles due to the restricted wavelength coverage of IRS. \label{OBIRsed.fig}}
\end{figure*}

\subsection{Detection of excess far-IR emission}

The photospheric flux of the sources is determined by taking a
black-body spectrum appropriate to the spectral type of the star,
giving a satisfactory fit to the 2MASS, WISE, and UBVRI photometry
accounting for distance and foreground extinction. { { In the
    optical region the reddening correction is important. We take for
    each sightline the extinction curve and foreground
    extinction as given by Valencic et al. (2004), Fitzpatrick (2004),
    Fitzpatrick \& Massa (2007), and Gordon et al. (2009). We compare
    this estimate of the photospheric flux with stellar atmosphere
    models taken from the Castelli \& Kurucz (2004) atlas for the
    stellar type given in Table~\ref{sample.tab}. The stellar
    atmosphere models are identical to black bodies at wavelengths $\geq
    0.8 \mu$m and in the optical continuum when reddened by additional
    0.1\,mag. }} Given the precision of IR photometry we believe the
uncertainties{ on} photospheric parameters to be less than
10\%. Temperatures for the black body spectra were based on the Heap
et al. (2006) calibration of O-star properties, and this was also used
as a starting point for the luminosities. Figure~\ref{OBIRsed.fig}
shows the IRS spectra and the contribution of the photospheres.  The
photospheric contribution thus determined was subtracted from the IRS
spectrum to identify excess emission. { {The precision of our
    estimate of the emission by the photosphere is exemplified in
    Fig.~\ref{pl_sedust.fig}.  Apparently our single stellar
    temperature fits the optical/IR photometry well. The model provides
    a reasonable estimate of the far-IR emission of the photosphere,
    even for multiple-component systems such as HD~149404.  }}

For 12 of the 22 sources, the fractional excess $(F_{\rm {IRS}} -
F_{\rm {bb}}) /F_{\rm {bb}} \ge 0.1$ over a large portion of the IRS
wavelength coverage, and for two stars, HD~149757 and HD~165024, the
detected excess is marginal (Fig.~\ref{OBIRsed.fig}). Ten of the stars
are{ not considered host excess emission}. Given the high
signal-to-noise ratio S/N of our IRS spectra with absolute flux
uncertainties of a few percent, this lower limit is somewhat
conservative but allows for reasonable (< 10\%) uncertainty in the
photospheric parameters. The excess becomes clearly visible at
wavelengths $\lambda \simgreat 10 \ \mu$m and in
Fig.~\ref{OBIRsed.fig} can be seen to be most pronounced at the
longest wavelengths, such that we refer to it as far--IR excess.


\begin{table*} [h!tb]
\begin{center}
  \caption{Spectral lines for stars with and without far--IR
    excess. We list for the optical high resolution spectra the name
    of instrument and date of the observations. Distance estimates are
    from GAIA, Ca~II and from the spectral and luminosity class
    (Sp/L).  The presence of the H~$\alpha$, C~III, and N~III lines is
    flagged. { For the Pfund~$\alpha$ ($n=6 \rightarrow 5$) and
      Humphrey~$\alpha$ ($n=7 \rightarrow 6$) hydrogen transitions
      observed with IRS we give the continuum subtracted line fluxes
       and equivalent width}. \label{hrs.tab}} \small
  \begin{tabular}{l lc rrr ccc  cccc }
    \hline
      \hline
      Star      &Instr.$^{\dag}$        & Date   & \multicolumn{3}{c}{Distance} &H~$\alpha$ & C~III     & N~III 
      & \multicolumn{2}{c}{Pf~$\alpha$}   & \multicolumn{2}{c}{Hu~$\alpha$ }\\
      \hline
      &                    &        & paral. & Ca~II & Sp/L       &${6563\,\AA}$&${5596\,\AA}$&${4641\,\AA}$ &7.5\,$\mu$m & EW & 12.3\,$\mu$m          & EW \\
      &                    &        & \multicolumn{3}{c}{(pc)}      &           &           &             & ($10^{-17}$\,W/m$^2$)  & ($\AA$) & ($10^{-17}$\,W/m$^2$) & ($\AA$)    \\
      \hline
          {\bf Excess}    &        &          &    &    &           &                   &         &       & &  \\
HD~24912   &   \scriptsize{GRANS}     &  \scriptsize{19960826}  & 725& 500& 640      & \scriptsize{no}                 & \scriptsize{yes}     & \scriptsize{yes}   & \small{$69 \pm 3.5  $}& 67& \small{$10.5 \pm 0.8  $}& 58 \\
HD~64760   &   \scriptsize{UVES}     &  \scriptsize{20090923}  & 360& 440& 525      & \scriptsize{weak wings}         & \scriptsize{no}      & \scriptsize{no}     & \small{$20 \pm 2.7  $}& 37 & \small{$3.9 \pm 0.7  $}& 42 \\
HD~149404  &   \scriptsize{UVES}     &  \scriptsize{20090308}  & 1315& 1865& 1240    &  \scriptsize{yes double}       & \scriptsize{yes}     & \scriptsize{yes}   & \small{$242 \pm 4.0 $}& 224 & \small{$41 \pm 1.0  $}& 176 \\
HD~149757  &   \scriptsize{HARPS}     &  \scriptsize{20060310}  & 112& 230 & 240   &  \scriptsize{weak$^b$}       & \scriptsize{no} & \scriptsize{no}               & \small{$<12.8       $}& - & \small{$<2.9          $}& - \\  
HD~151804  &   \scriptsize{FEROS}    &  \scriptsize{20040502}  & 1640& 1680& 1365   &  \scriptsize{yes + H$\beta$}   & \scriptsize{yes}     & \scriptsize{yes}    & \small{$307 \pm 3.3 $}& 473 & \small{$39 \pm 0.8  $}& 254 \\ 
HD~154368  &   \scriptsize{UVES}     &  \scriptsize{20090306}  & 1220& 1290& 1290    &  \scriptsize{marginal}           & \scriptsize{yes}     & \scriptsize{no}    & \small{$39 \pm 2.8  $}& 81 & \small{$6.5 \pm 0.7  $}& 72 \\ 
HD~165024  &   \scriptsize{FEROS}    &  \scriptsize{20040706}  & 280&  440&  350    & \scriptsize{no}                 & \scriptsize{no}      & \scriptsize{no}     & \small{$<11         $}& - & \small{$<2.6          $}& - \\  
HD~188209  &   \scriptsize{BOES}     &  \scriptsize{20050526}  & 1490& 1410& 1270    & \scriptsize{yes}                & \scriptsize{yes}     & \scriptsize{no}    & \small{$9.7 \pm 1.9 $}& 57 & \small{$2.6 \pm 0.5  $}& 84 \\
HD~190918$^a$&   \scriptsize{BOES}     &  \scriptsize{20070615}  & 1960&  3440& 2290  & \scriptsize{yes}                & \scriptsize{yes}     & \scriptsize{?}   & \small{$60  \pm 2.9 $}& 274 & \small{$9.4 \pm 0.8 $}& 174 \\  
HD~192639$^a$&   \scriptsize{ELODIE}   &  \scriptsize{{ 20010811}}   & 2600& 2080&  2115  &  \scriptsize{\scriptsize{yes} asym} & \scriptsize{yes}    & \scriptsize{yes}   & \small{$32 \pm 1.9   $}& 221 & \small{$3.0 \pm 0.5 $}& 96 \\  
HD~206773$^a$&   ---      &  ---       & 960& 550& 595      & ---                &         & ---  & \small{$41 \pm 2.2  $}& 154 & \small{$6.8 \pm 0.6 $}& 117 \\  
HD~210839  &   \scriptsize{MAESTRO}  &  \scriptsize{20040828}  & 620&  1250& 1190   & \scriptsize{yes}                & \scriptsize{yes}     & \scriptsize{yes}    & \small{$134 \pm 3.2 $}& 187 & \small{$19 \pm 0.8 $}& 129 \\  
          \hline
{\bf No excess} &        &            & &&                   &                &         &       & &    \\
  HD~34816   &  \scriptsize{Sandiford} &  \scriptsize{19930210}  & 270& 410&  380     & \scriptsize{no}                 & \scriptsize{no}      & \scriptsize{?}    &\small{$<7.2         $}& - & \small{$<1.6 $}& - \\
HD~36861   &  \scriptsize{UVES}      &  \scriptsize{20040205}  &340& 530& 490      & \scriptsize{no}                 & \scriptsize{yes}     & \scriptsize{yes}     &\small{$11.3 \pm 3.4 $}& 16 & \small{$ <2.4$}& - \\
HD~38087   &  \scriptsize{UVES}      &  \scriptsize{20050321}  & 340& 470& 490      & \scriptsize{no}                 & \scriptsize{?}       & \scriptsize{no}     &\small{$ 5.9 \pm 1.6 $}& 121 & \small{$ <1.4 $}& - \\
HD~47839$^b$&  \scriptsize{GRAMS}    &  \scriptsize{19960405}  & 280& 710& 665      & \scriptsize{no}               & \scriptsize{yes}     & \scriptsize{yes}     &\small{$ <6.3        $}& - & \small{$<1.7 $}& - \\
HD~53367   &  \scriptsize{UVES}      &  \scriptsize{20091215}  & --& 1000& 780      & \scriptsize{yes}        & \scriptsize{?}       & \scriptsize{yes}            &\small{$8.8  \pm 0.8 $}& 39 & \small{$  1.1 \pm  0.2 $}& 33 \\
HD~62542$^a$   &  \scriptsize{UVES}      &  \scriptsize{20031113}  & 390& 450& 530      & \scriptsize{no}                 & \scriptsize{no}      & \scriptsize{no} &\small{$0.14 \pm 0.04$}& 6 & \small{$  1.1\pm  0.2$}& 33 \\
HD~65575$^a$  &  ---       &  ---       & 420& 155& 390      & ---                & ---     & --- &\small{$<10.0        $}& - & \small{$< 2.3 $}& - \\
HD~108639$^a$&  ---       &  ---       & 1820& 2380& 2280    & ---                & ---     & --- &\small{$0.6 \pm 0.1  $}& 16 & \small{$ 0.19 \pm 0.03 $}& 37 \\
HD~147933  &  \scriptsize{HARPS}     &  \scriptsize{20070330}  & 145& ---& 100      & \scriptsize{no}                 & \scriptsize{no}      & \scriptsize{no}     &\small{$<  12.0      $}& - & \small{$<2.7 $}& - \\
HD~214680  &  \scriptsize{MAESTRO}   &  \scriptsize{19660704}  & 360& 690& 615      & \scriptsize{no}                 & \scriptsize{no}      & \scriptsize{no}     &\small{$<   5.9      $}& - & \small{$<1.5 $}& - \\
\hline
\end{tabular}
\end{center}
$^{\dag}$ Optical high-resolution spectra of this instrument are made
available at CDS.  $^a$ Parallax from GAIA (DR2) by the Gaia
Collaboration (Brown et al. 2018) $^b$ Out of five spectra taken
between 1993 -2010 only the HARPS spectrum taken 2006-03-10 shows weak
H~$\alpha$ emission (Fig.~\ref{Halpha149757.fig}).
\end{table*}

\subsection{High-resolution spectroscopy}

For our sample, we collate high-resolution optical spectra observed
over the past 25 years. We have used ESO telescopes with instruments
UVES/VLT offering a resolving power of $R =\lambda / \Delta \lambda
\sim 80,000-110,000$, HARPS ($R = 115,000$) at the 3.6\,m, and FEROS
($R = 48,000$) at the ESO/MPG 2.2\,m. We also used spectrographs at
other observatories such as ELODIE ($R = 42,000$) at the 1.93\,m
Observatoire de Haute-Provence, GRAMS ($R = 40,000$) at the 1\,m of
the Russian Special Astrophysical Observatory at Northern Caucasus,
Sandiford ($R = 64,000$) at the 2.1\,m Otto Struve telescope, BOES ($R
= 30,000 - 90,000$) at the 1.8\,m Bohyunsan Optical Astronomy
Observatory in Korea, and MAESTRO ($R = 40,000-120,000$) at the 2\,m telescope of
the Terskol Observatory at Northern Caucasus.

In Table~\ref{hrs.tab} we list for each star the spectrograph used
together with the observing date. We derive distance to the stars from
GAIA DR2 parallaxes (Gaia Collaboration, Brown et al., 2018), Ca~II
(Megier et al. 2009), and from the spectral and luminosity class
(Sp/L). We note that there is often a large scatter of up to a factor of
two in the distance estimates (Kre{\l}owski 2018). We comment on the
presence or absence of the H~$\alpha$ at ${6563 \ \AA}$, C~III at
${5596 \ \AA}$, and N~III at ${4641\ \AA}$ lines and specify the
strength of the Pfund~$\alpha$ ($n=6 \rightarrow 5$) near $7.5\ \mu$m
and Humphrey~$\alpha$ ($n=7 \rightarrow 6$) transitions near $12.3
\ \mu$m (Kramida et al. 2018), as measured in the Spitzer/IRS
spectrum.  { Pfund and Humphrey, C~III and N~III lines are detected
  in emission.}  Spectral classification of the stars is based on{
  our optical HR spectra} and reported in Table~\ref{sample.tab}. Our
classifications generally agree with those derived by Bowen et
al. (2008), Jenkins et al. (2009), Rauw et al. (2015), Skiff (2013),
Sota et al. (2011), and Sota et al. (2014).

\begin{figure} [h!tb]
\includegraphics[width=10.4cm,clip=true,trim=1.5cm 0.cm 0cm 0.cm]{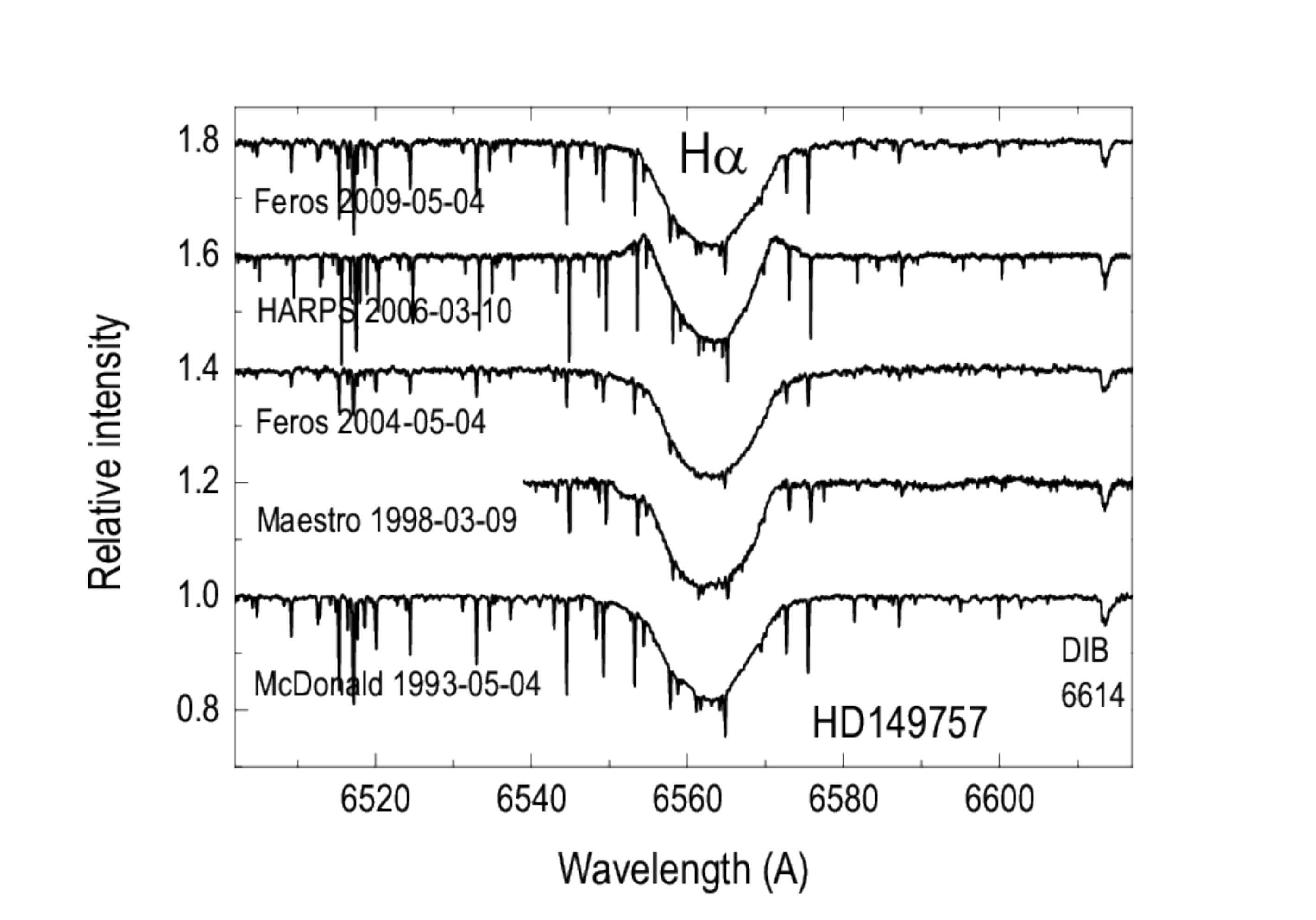}
\caption{High-resolution spectra of HD~149757 between observing epochs
  from 1993 and 2009. H~$\alpha$ emission is detected only in the 2006
  spectrum; we note the wings of the absorption profile. \label{Halpha149757.fig}}
\end{figure}

Free-free radiation is emitted wherever there is ionised gas. The {
  presence} of hydrogen lines can also be a transient phenomenon and
prototypical examples are Be-stars. The transient nature of the
H~$\alpha$ emission in our sample is best demonstrated by HD~149757.
The star is a rapid rotator and frequently classified as O9.5. We
confirm the more recent O9.2IV classification by Sota et al. (2014),
whereas Levenhagen \& Leister (2006) classified the star as Be-star
(type B0Ve).  Their classification is based on a spectrum taken April
2000 at the ESO 1.2m.  We have taken multiple high resolution spectra
between May 1993 and March 2010 (Fig.~\ref{Halpha149757.fig}). With
the exception of a spectrum taken in 2006 where we observed weak
H~$\alpha$ emission in the wings of the absorption line, H~$\alpha$
emission is not detected in any of these spectra of HD~149757. The
Spitzer IRS spectrum taken in September 2008 displays Pfund~$\alpha$
and Humphrey~$\alpha$ in emission.

\begin{figure} [h!tb]
  \includegraphics[width=7.cm,angle=90.,clip=true,trim=0cm 0.cm 0cm 3.cm]{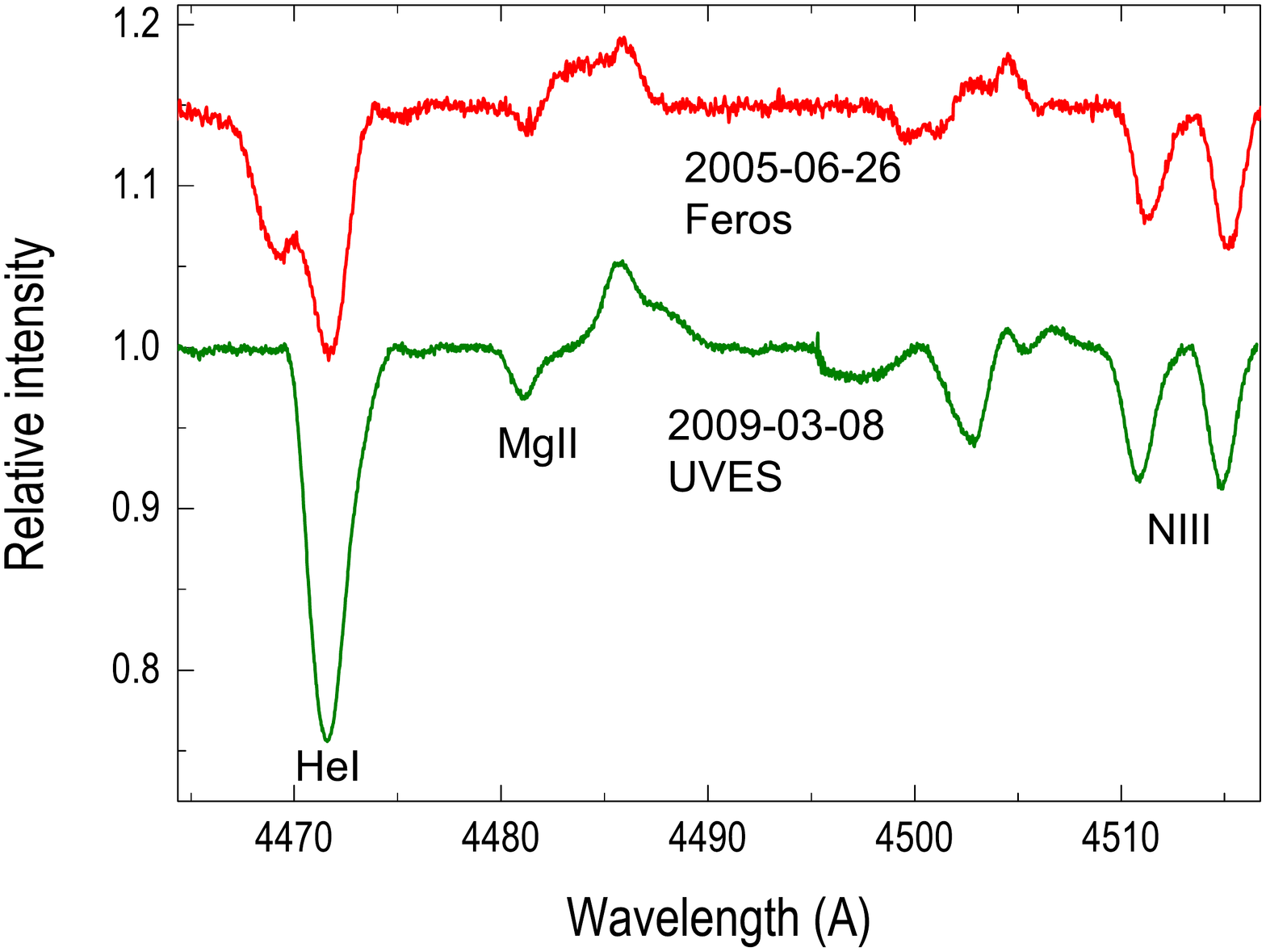}
  \includegraphics[width=7.cm,angle=90.,clip=true,trim=0cm 0.cm 0cm 3.cm]{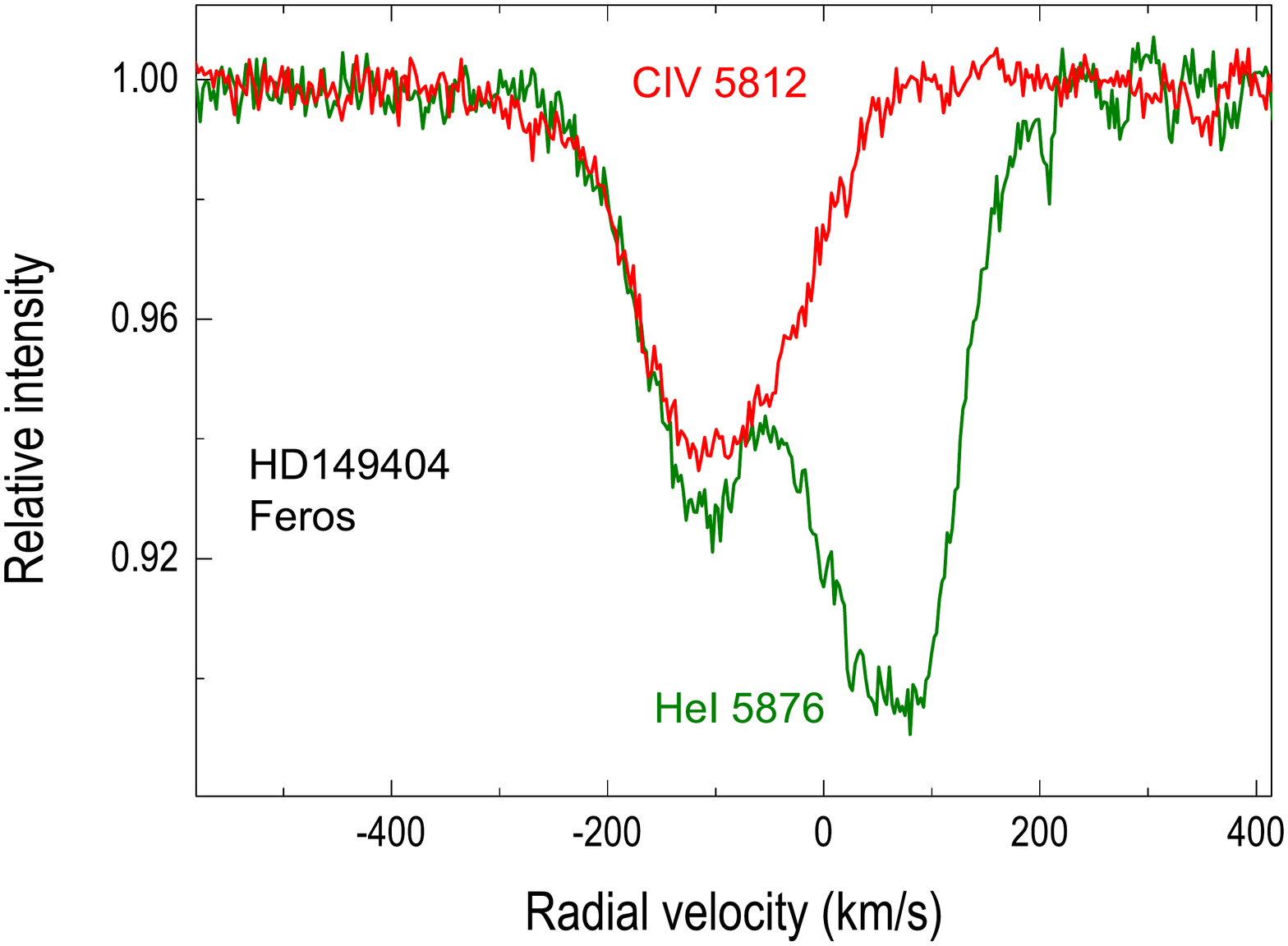}
  \caption{High-resolution spectra (top) of HD~149404 taken with FEROS
    in 2005 (red) and UVES in 2009 (green).  The lower panel shows the
    velocity dispersion. One notices an additional component observed in the
    HeI 5876~\AA\ line, likely originated in a  companion
    star, while in C~IV such a second component is absent.
 \label{HD149404.fig}}
\end{figure}

We present a second example{ underscoring} the transient nature of
the lines and show high-resolution FEROS and UVES spectra of HD~149404
in Fig.~\ref{HD149404.fig}.  The He~I { lines} at 4471 and 5876 \AA
\/ observed with FEROS demonstrate that HD~149404 is a binary.  One of
the components shows the C~IV { 5812 \AA} line and is therefore an O
type star while the second does not show the C~IV { 5812 \AA} line
and is a B type star. The combined spectrum is apparently variable.


We detect H~$\alpha$ emission in $\sim 60\%$ (7 out of 12) of the
far-IR excess stars and in only 14\% (1 out of 7) of the { non-}far-IR
excess stars. Therefore we cannot firmly conclude from optical {
  recombination} lines whether a far-IR excess is present or not.
Interestingly, the  { non-}far-IR excess star HD~47839 displays C~III
and N~III but no hydrogen lines despite it being classified as a
Be-star.

Spitzer/IRS offers the unique capability of observing hydrogen lines
simultaneously with the far-IR emission.  This is of great advantage
because of the transient nature of the lines. { The strongest H
  recombination lines observed in the IRS domain} are the
Pfund~$\alpha$ ($n=6 \rightarrow 5$) and Humphrey~$\alpha$ ($n=7
\rightarrow 6$) transitions near $7.5\ \mu$m and $12.3 \ \mu$m,
respectively. { Pf~$\alpha$ lines are stronger than Hu~$\alpha$
  lines}. We find that there is { no strong correlation} between
the detection of these lines and the presence of a far-IR
excess. Although from inspection of Table~\ref{hrs.tab} a clear trend
can be reported: the line intensities are typically an order of
magnitude stronger in{ stars with far-IR excess than in stars
  without far-IR excess.} The detection rate of the lines is $\sim
84$\% (10 out of 12) for IR-excess stars whereas Pf~$\alpha$ is
detected only in 50\% and Hu~$\alpha$ in 33\% for stars without IR
excess.


\subsection{High-contrast imaging}

The IRS spectra reveal a high detection rate of excess far-IR emission
over the predicted photospheric component for 12 out of the 22 massive
stars (55\%). We aim to resolve a dust component in a possible
disk-like structure in scattered starlight using SPHERE, the
extreme-adaptive-optics instrument at the VLT (Beuzit et
al. 2008). The detection of scattered-light disks around massive stars
would have presented interesting historical parallels to the detection
of dusty disks around low-mass stars, such as $\beta$~Pic, which was
first identified in the IR using IRAS before being confirmed by
scattered-light imaging (Smith \& Terrile 1984).

We performed near--IR (H band) high-contrast coronagraphic imaging
observations of three stars that show excess far-IR emission:
HD~149404, HD~151804, and HD~154368, together with { HD~147225 as a reference
  star}. We used the IRDIS sub-instrument of SPHERE (Beuzit et
al. 2008) on April 19 and June 19, 2015. { The targets were
  observed for approximately 45 minutes in the broad H band filter
  with a detector integration time of 0.84\,s and a pixel scale of
  12.27\,mas.} All exposures are flat-field corrected and background
subtracted, their bad pixels are removed, and the detector dithering is
corrected. Before adding the exposures, a subpixel shift algorithm is
used to centre the images over time, and then the { IRDIS detector
  channels} are averaged. The procedure is performed on the science
and reference star. The latter was used for point spread function
(PSF) subtraction applying the algorithm described by Lafreniere et
al. (2008). Finally, we compute the contrast given as the flux ratio
$F(r)/F_{*}$ of science target and stellar halo as estimated from the
PSF. The contrast is a function of distance $r$ (separation) from the
star.

\begin{figure} [h!tb]
\includegraphics[width=9.5cm,clip=true,trim=2cm 2.cm 0.5cm 3cm]{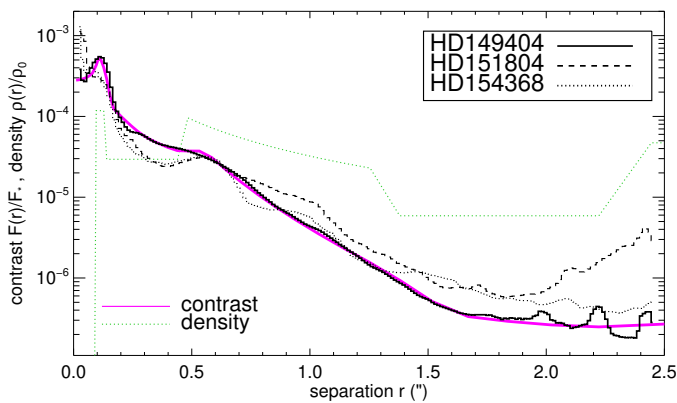}
\caption{SPHERE/IRDIS upper limits ($3\sigma$) of the contrast
  $F(r)/F_{*}$ (black) as a function of separation $r$ from the
  stars. A dust model fitting the upper limit of the contrast of
  HD~149404 (magenta line) is shown together with the applied dust
  density distribution $\propto \rho(r)/\rho_{0}$ (green dotted
  line). \label{pl_contrast.fig}}
\end{figure}

We do not observe any significant structures above the noise of the
stellar halo that one could claim as scattered-light disk detection.
A 1$\sigma$ noise limit of the contrast curve is computed.  It is
given for each pixel separation as standard deviation in annuli with
one-pixel width centred on the PSF core.  Further details of the
observing set-up and data reduction are given by Banas (2017) and
Scicluna et al. (2017).  The upper limits of the contrast curves of
the three stars are displayed in Fig.~\ref{pl_contrast.fig}.  { At $\sim 2\arcsec$ separation from the star, our data reaches a 1$\sigma$
contrast of $2 \times 10^{-7}$,} demonstrating the unparalleled imaging
contrast that can be achieved with SPHERE/IRDIS.

\section{Discussion}

After the detection of the far-IR excess in{ 12 out of 22} OB stars,
we{ explore} the possible origin; we discuss the impact of background
sources, the Be phenomenon, and emission from a faint dust halo around
the sources. {Spectral energy distribution (SED) models} of the
stars are best constrained when Spitzer spectra are available together
with Herschel photometry. This is the case for five stars with far-IR
excess that are listed in Table~\ref{model.tab} (column~1) together
with their stellar temperature (column~2), luminosity (column~3), and
mass (column~4). To quantify the strength of the excess far-IR
emission, we compute{ the ratio of the observed and photospheric
  luminosity in the IRS $5 - 35 \ \mu$m spectral range} (column~5).

\subsection{Background source}

To identify possible contamination by dusty background sources, we fit
a single-temperature modified black body, $f_{\nu} \propto \nu ^{2}
B_{\nu}(T)$, to the continuum excess using MPFIT (Markwardt 2009). The
reduced-$\chi^2$ values are all $\gg 1$, indicating that a single
black body is not adequate for fitting these spectra. Given the high
temperatures of the best fits (150-290\,K) it is also clear that a
cool background source is not able to explain the excess.

\begin{table*} [h!tb]
\begin{center}
  \caption{Characteristics of far-IR excess sources with available
    Spitzer/IRS and Herschel photometry. \label{model.tab}}
  \begin{tabular}{l ccc ccc rrc cc}
    \hline
    \hline
  1& 2&  3& 4& 5& 6& 7& 8& 9& 10 & 11 & 12\\
  \hline
  Source   &  \multicolumn{3}{c}{Photosphere}  &  & & & \multicolumn{3}{c}{Free-free emission}
  & \multicolumn{2}{c}{Dust shell} \\ 
  \hline
  & $T_{\rm{eff}}$ & $\log{\left(\frac{L_{\rm {*}}}{L_{\odot}}\right)}^a$  & $M_{\rm {*}}$ & 
  $\frac{L_{\rm {IRS}}}{L{'}_{*}}$  &
  $\dot{M}$ &  $\dot{M}_{\rm {ff}}$ & $F^{\rm {ff}}$ & $F^{\rm {wind}}$ & $F^{\rm {obs}}$ & $A_{\rm{V, CS}}^b$ & $M_{\rm {dust}}^c$ \\
  &(K) & &  $\left(M_{\odot}\right)$ & & \multicolumn{2}{c}{$\left(\log\frac{M_{\odot}}{yr}\right)$} &
  \multicolumn{3}{c}{3.6\,cm flux $\left(\mu\rm{Jy}\right)$ }  & ($\mu$mag) &($\mu$\Msun )\\
  \hline
HD~24912        &38000 &5.3& 31 & 1.2&  -6.2 &-5.5 &258& $30 $ & $<120^d$&  95  & 62 \\
HD~149404       &33000 &5.7& 19 & 1.8&  -- & -- &1533&$ 300$ & –       &  140 & 92 \\
HD~149757       &33000 &4.7& 20 & 1.1&  -7.4 &-6.6 &116 &$ 9$   & –       &  95  & 62 \\
HD~151804       &35800 &5.8& 22 & 1.4&  -5.0 &-4.6 &987 &$ 300$ &$ 400^e$ &  95  & 62 \\
HD~210839       &41000 &6.0& 21 & 2.1&  -4.7 &-4.7 &602 & $600$ & $430^d$ &  110 & 80\\
\hline
\end{tabular}
\end{center}
$^a$Applying Sp/L distance as of Table~\ref{hrs.tab}. $^b$Visual
extinction from inner to outer radius, and $^c$total dust mass. $^d$Puls
et al. (2006), $^e$Lamers \& Leitherer (1993).

\end{table*}

\subsection{Be-stars}

Be-stars show a B-type stellar spectrum in combination with Balmer
line emission, and an IR excess is often observed. The IR excess of Be
stars is due to free-free emission from an ionized purely gaseous
circumstellar disk and does not come from a wind or synchrotron
component (Rivinius et al. 2013). For dense disks, the excess starts to
dominate the photospheric emission at near- to mid-IR wavelengths
(Vieira et al. 2017). The slope of the IR excess depends on the
density profile of the disk, and is close to a power-law
$\nu^{\gamma}$ with $0.6 \simless \gamma \simless 2$ (Klement et
al. 2017). The Be phenomenon may be a period in the life of{ a}
`normal' B star (Galazutdinov \& Kre{\l}owski 2006).

We searched various catalogues and publications listing Be-stars
(Fabregat et al. 1996, Rivinius et al. 2006, Wisniewski et al. 2007,
Catanzaro 2013, Draper et al. 2014, Chojnowski et al. 2015, Lin et
al. 2015) and find three stars in our sample that were classified as
Be-type stars. There is HD~47839 that does not display a far-IR
excess, HD~149757 ($\zeta$~Oph) that occasionally shows Be-type
outbursts (Vogt \& Penrod 1983,  Kambe et al. 1997,
  Fig.~\ref{HD149404.fig}), and HD~206773 that has a strong far-IR
excess with a rising wavelength dependency at $\simgreat 25 \ \mu$m
(Fig.~\ref{OBIRsed.fig}).  The occurrence of three likely Be-stars in
our sample of 22 stars is in line with the estimate that about 17\% of
B-type stars are Be-stars (Zorec \& Briot 1997).  However, we detect a
far-IR excess in 12 out of 22 sources at a much higher detection
rate. Fully ionized optically thick disk models of Be-stars are
presented by Carciofi \& Bjorkman (2006).  They predict excess
emission for disks inclinations $\le 60^{\degr}$ emerging at short
wavelengths ($\simless 1 \ \mu$m) and for edge-on view ($90^{\degr}$)
in the mid-IR at $\sim 10 \ \mu$m. The excess that we are observing
develops in the far-IR at wavelengths $\simgreat 10 \ \mu$m
(Fig.~~\ref{OBIRsed.fig}). It is unlikely that all our far-IR excess
stars are yet unclassified Be-stars with their disks viewed nearly
edge-on. Therefore the Be-star phenomenon is unlikely to be the
dominant origin of the far-IR excess.

\subsection{Wind models}

It has been shown (Barlow 1979) that a dense ionised medium generates
free-free emission at wavelengths from radio to IR. A simple
$F_{\rm{ff}} \propto \nu^{0.7}$ dependence has been shown to fit well
over{ six orders of magnitude in frequency} from the IR to the
radio (Klement et al. 2017). { It} is apparent for the strongest
radio-emitters (e.g. HD~210839) that an IR free-free component is
present. In order to investigate this effect we located relevant radio
data for our targets where available. For those stars with no radio
data, it is possible to estimate a theoretical mass-loss rate
$\dot{M}$ (Vink et al. 2001), as a function of the mass, luminosity,
temperature and metallicity of the source\footnote{using a routine
  from http://star.arm.ac.uk/~jsv/Mdot.pro.}. As a result, it is
necessary to determine masses for our sample of stars. To do so, we
compare the luminosity and temperature to the Geneva evolutionary
tracks (Ekstr\"om et al. 2012) for solar metallicity to find a best
fitting mass. Multiple systems should be excluded from such an
analysis. The mass-loss rates thus-derived are then used to predict a
3.6\,cm flux for the wind $F^{\rm{wind}}$. The flux estimate is based
on

\begin{equation} \label{Mdot.eq}
  \dot{M} = 5.32 \times 10^{-4} \ \frac{\mu
    D^{\frac{3}{4}}\ \rm{v}_{\infty}}{\sqrt{\tilde{g} \ \nu}}
  \ \left(F^{\rm{wind}}_{\nu} \right)^{\frac{3}{4}} \ \left(
  \frac{M_{\odot}}{yr} \right) \,
,\end{equation}
 
where { $D$ is the distance}, v$_{\infty}$ the wind thermal velocity,
and $\tilde{g}$ is the Gaunt factor assuming an effective wind
temperature $T_{\rm {wind}} \sim 0.6 \ T_{*}$ (Carciofi \& Bjorkman
2006). { {In the model by Vink et al. (2001) it is assumed}} that the
wind is composed entirely of ionised hydrogen, and so the mean ionic
weight $\mu = 0.5$ in units of proton mass. The derived mass-loss
rates are given in Table~\ref{model.tab} (column~6) together with the
3.6\,cm flux $F_{\rm {wind}}$ (column~8).  We extrapolate the fit to the
IR data by the free-free component to a 3.6\,cm flux $F_{\rm{ff}}$
(column~7), which using Eq.~\ref{Mdot.eq} gives a mass-loss rate
$\dot{M}_{\rm {ff}}$ specified in column 7. Whenever available we give
the observed 3.6\,cm flux $F_{\rm{obs}}$ in column 9.

In Fig.~\ref{pl_sedust.fig} (left panels) the free-free component
$F^{\rm {ff}}$ is shown using a blue dotted line, and the total flux $F_{*}
+F_{\rm{ff}}$ using a green full line. We find that the free-free component
can always be adjusted so that IR data are matched.  The SEDs of
HD~149757 and HD~210839 are well fit when assuming that the free-free
flux is given by the wind models ($F_{\rm{ff}}=F_{\rm {wind}}$) while
such a model underestimates IR data of the other sources. However, the
required 3.6\,cm free-free flux is much stronger (factor $\sim 2$)
than the observed fluxes of HD~24912 and HD~210839 by Puls et
al. (2006) and HD~151804 by Lamers \& Leitherer (1993).


\subsection{Circumstellar dust models}

As a single dust temperature cannot reproduce the observed emission,
we require a distribution of dust near the source with a range of
temperatures and masses: a circumstellar envelope. The data for each
source were modelled assuming spherical symmetry using a code described
in Kr\"ugel (2008) with dust heated by a central star. The code solves
the radiative transfer equation by ray tracing, with the source placed
at the centre of the cloud. We solve the radiative transfer equation

\begin {equation}
 I_{\nu}(\tau) =  I_{\nu}(0) \ e^{- \tau} \ + \int_0^{\tau} S(x) \ e^{x - \tau} \ dx 
,\end {equation}

with the source function

\begin {equation}
 S_{\nu} = \frac{K^{\rm{abs}}_{\nu} \ \int{P(T) \ B_{\nu}(T)\ dT} + (1 - g_{\nu})\ K^{\rm{sca}}_{\nu} \ J_{\nu} }{K^{\rm{ext}}_{\nu}} \, 
,\end {equation}

where $I_{\nu}$ is the intensity, $B_{\nu}(T)$ is the Planck function,
$P(T)$ is the temperature distribution for stochastically heated
grains (Siebenmorgen et al. 1992), $K^{\rm{abs}}$ is the absorption,
$K^{\rm{sca}}$ the scattering, $K^{\rm{ext}}$ the extinction
cross-sections, $g$ is the scattering anisotropy
parameter, and $\tau$ is the optical depth. By multiplying the
scattering cross-section by $1 - g$ we are able to account for
anisotropic scattering by effectively dividing the scattering into a
purely isotropic component and a forward scattered component (Scicluna
\& Siebenmorgen 2015). The boundary conditions are provided by the
external radiation field and the stellar flux. The input stellar
spectrum is treated as a black body of given temperature and
integrated luminosity truncated at Ly~$\alpha$, which is then processed
by the dust distribution.  It is not necessary to treat an
interstellar radiation field as the luminosity of the stars is so
great that they dominate the local radiation field (Mathis et
al. 1983). Kr\"ugel (2008) gives the interstellar
radiation field a mean value of 0.04\,erg~s$^{-1}$~cm$^{-2}$, while for a star of luminosity
$10^4$\,\Lsun \/ the radiation field at 1\,pc is 0.32\,erg~s$^{-1}$~cm$^{-2}$, an
order of magnitude larger. The orbital separations of the multiple
stars included in the sample are small compared to the expected dust
inner radius, so we model them as a single point source.

\begin{figure*} [htb]
\includegraphics[width=16.cm,clip=true,trim=0.3cm 1.cm 1.2cm 1.2cm]{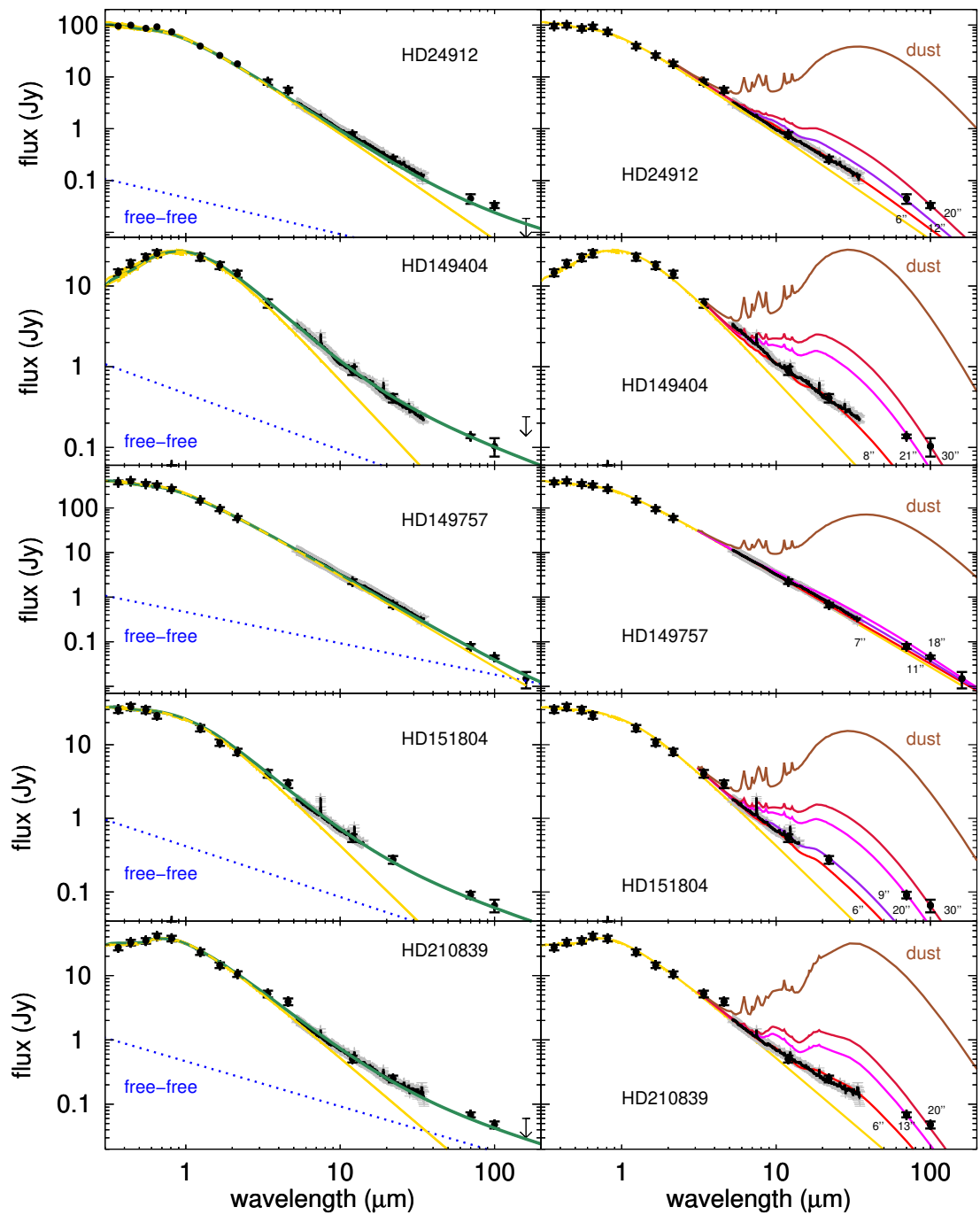}
\caption{SED of massive stars. The total emission is fit by adding to
  the photosphere (yellow) either free-free (left blue dotted lines) or
  dust emission in apertures as labelled (right red-brown lines). Data
  (black) are from optical catalogues as available at VizieR/CDS,
  2MASS, WISE, Spitzer/IRS, and Herschel/PACS with 1$\sigma$ error
  bars and 3$\sigma$ upper limits. \label{pl_sedust.fig}}
\end{figure*}

The dust model consists of sub-micron sized grains of amorphous carbon
and/or silicates, as well as stochastically heated particles in the
form of PAHs and nanometre-scale graphite { and silicate} particles
(Siebenmorgen et al. 2014). { We apply dust parameters for
  single-cloud sightlines given by Siebenmorgen et al. (2018). } The
radiative transfer is solved iteratively, and dust emission and
temperatures are solved self-consistently. The geometry of the
envelopes is characterised by an inner and outer dust radius, and a
dust density distribution $\rho(r)$.  The model allows the computation
of SEDs within apertures of the
circumstellar envelope of different angular sizes.

We demonstrate the models using the strong far-IR excess source HD~149404
for which a SPHERE contrast curve and Herschel photometry are
available. Its photosphere is approximated by a single temperature of
33,000\,K, which is reddened by $A_{\rm {V}} = 2.2$\,mag, and a
Fitzpatrick (2004) and Gordon et al. (2009) extinction curve with $R_{\rm
  {V}} = 3.65$. As shown in Fig.~\ref{pl_sedust.fig}, the photosphere
model fits the optical (UBVR), 2MASS, and WISE photometry despite the
variability of this multi-component system
(Fig.~\ref{HD149404.fig}). The lower panel of Fig.~\ref{HD149404.fig}
shows an additional component in the HeI~5876\AA\ line representing
the fainter O-type companion star, while{ such a second component
  is absent in the C~IV 5812 \AA\ line.} WISE photometry at 12 and
22\,$\mu$m is consistent with the Spitzer IRS spectrum.

We take the SPHERE contrast curve to constrain the innermost
$0.05\arcsec - 2.5\arcsec$ region of the dust shell
(Fig.~\ref{pl_contrast.fig}).  The IR data are fit assuming a constant
dust density in a shallow dust envelope in the range $0.011 < r \leq
1$\,(pc) of $A_{\rm V} = 300 \ \mu$mag.  We also compute a second
model in which the dust content in the inner region ($\leq 2275$\,AU)
is increased by multi-component power laws until the $3\sigma$ upper
limit of the contrast curve is matched. This gives an upper limit of
the total amount of warm ($\simgreat 250$\,K) dust of $\simless
10^{-10}$\,\Msun \/.  The density distribution in the inner region is
shown as green and the contrast curve of that model is shown with a
magenta line in Fig.~\ref{pl_contrast.fig}. The low amount of warm
dust does not contribute to the SED in the IRS spectral range. { For HD~24912, HD~149757, and HD~151804, we  keep} the same structure of
the circumstellar dust shell and set $\rho(r) =\rho_0$ to be constant
for $0.011 < r \leq 1$\,(pc) and otherwise there is no dust. { For
  HD~210839, our highest-luminosity far-IR excess star, we increased
  the inner radius to 0.026\,pc so that the model does not show the
  SiO dust bands. At the inner radius, the dust temperature is about
  300\,K. Nano-sized particles are photo-dissociated up to distances
  of $\sim 6 \times 10^{16}$\,cm from the star. The models are
  consistent with the non-detection of PAH features or oxygen-rich
  dust bands (silicates) in the IRS spectra. The dust emission
  spectrum is insensitive to the adopted effective temperature of the
  OB stars.  The computed dust emission spectra are identical when
  stars of the same luminosity and of spectral shapes ranging between 30,000
  and- 40,000\,K are considered.  Therefore,} we only allow  $\rho_0$ to vary as a free parameter when fitting the far-IR data. The corresponding extinction
$A_{\rm {V, CS}}$ and dust mass $M_D$ are listed in
Table~\ref{model.tab}. Spectral energy distribution models are shown for different observing
apertures of the circumstellar dust shells in the right panels of
Fig.~\ref{pl_sedust.fig}.{ A similar-quality fit is achieved for the SEDs using  either the free-free or dust model, and so a preference for one or the other model was not decided upon.  At the spatial resolution of ALMA
  the extended dust halo model predicts submillimetre fluxes that are
  orders of magnitude fainter than the point-like emission that is
  estimated by the free-free model.  Therefore the nature of the
  far-IR excess can be tested by future ALMA observations in the
  submillimetre continuum. A similar argument holds for observations in
  the radio, however there is the underlying assumption that one can
  extrapolate the power law of the free-free component over many
  orders of magnitude in frequency.}


\subsection{Dust formation}

The observed far-IR excess may be due to free-free or dust emission,
and for the latter, one wonders how grains may sustain the harsh
environment of the OB stars. For A stars, the blow-out radius or the
dust grain radius below which all grains experience greater radiation
pressure forces than gravitational is several micrometres (Beust
2010). The blow-out radius is proportional to $L_*/M_*$ implying that
it is between 1 and 2 orders of magnitude larger for OB stars. This
means that sub-micrometre sized grains should be rapidly removed and
only much larger particles would remain in the stellar environment.
Another important effect to consider is Poynting-Robertson drag, which
causes dust grains to lose angular momentum and spiral toward the star
(Draine 2011). Once again, this effect is stronger for smaller grains
around more luminous stars and would cause grains between 0.1 and
1\,mm to spiral in from the inner radius of our models on timescales
of $10^6 - 10^7$\,yr. Therefore, this dust population is not
long-lived and must be replenished from some source. A replenishment
similar to collisional cascades observed by debris-disk emission (Su
et al. 2013) is unlikely because of the { non-detection} of a
scattered-light disk in the high-contrast imaging observations with
SPHERE of at least three of our far-IR excess stars.

\begin{figure} [h!tb]
\includegraphics[width=9cm,clip=true,trim=3cm 0.cm 3cm 0.cm]{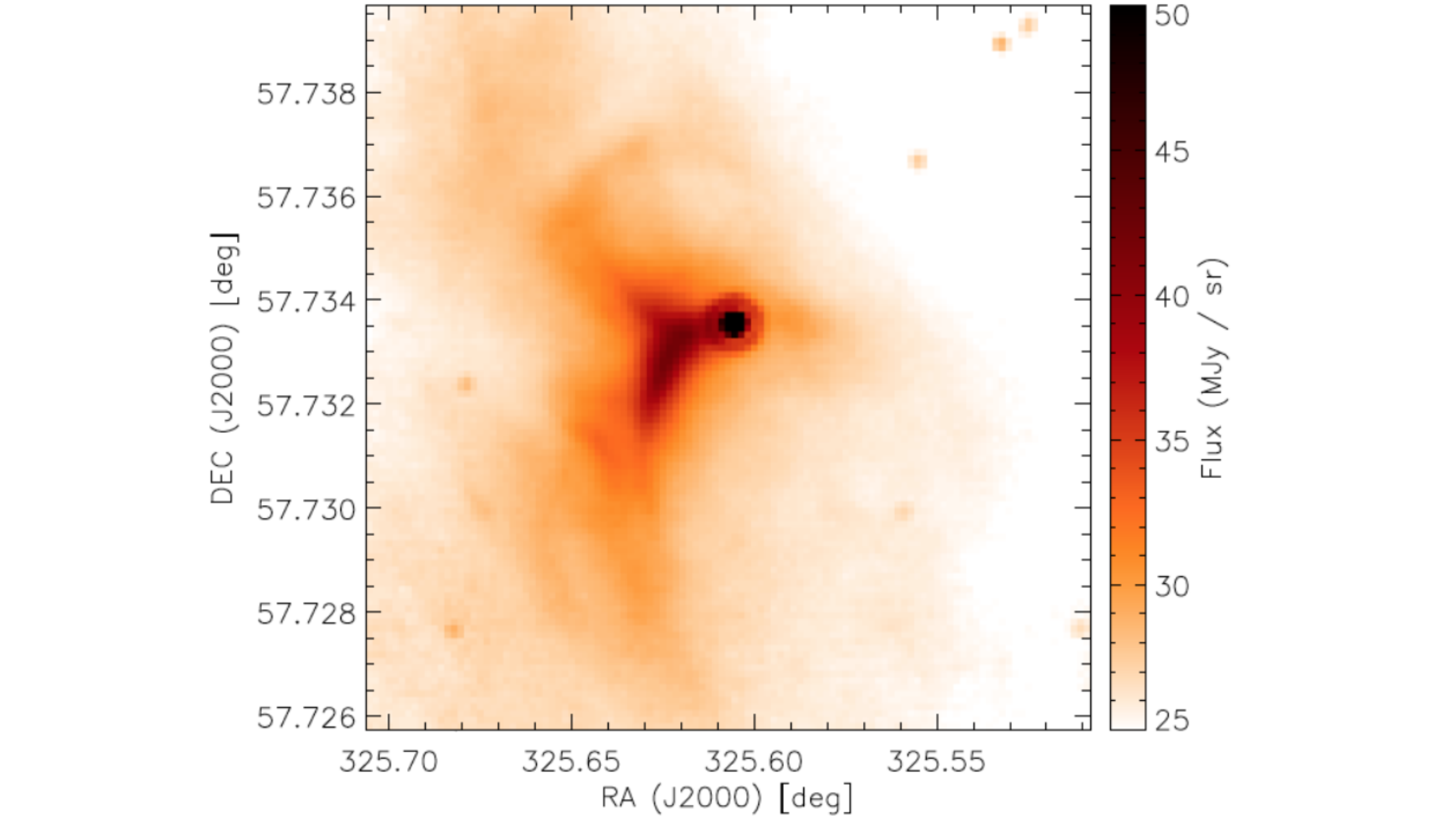}
\caption{Spitzer/MIPS 24\,$\mu$m image of HD~206773. Right ascension and
  declination is in degrees, the colour bar shows flux in MJy/sr. We note
  the large-scale background structure to the west. \label{ima24.fig}}
\end{figure}

One possible cause for stars to have both dust and wind emission is
that dust formation takes place in the winds. This is most likely
occurring in multiple-systems where the wind collisional region
between the two components is likely to reach extremely high
densities, allowing the condensation of dust particles, as in, for
example, the WC7+O5 binary WR140 (Williams 2008). From our data set,
we cannot rule out the presence of lower-luminosity companions that
are also able to drive a wind, particularly for the most luminous
stars.  A further suggestion for those stars that have left the main
sequence is that they could have undergone phases - for example, as a
red supergiant - where a cooler photosphere may have driven a dusty
wind which is now being cleared.  If either of these scenarios is
accurate, it would provide an efficient way for a stellar population
to produce large quantities of dust before the appearance of red
supergiants, supernovae, and AGB stars, which may help to explain the
observations of anomalously high dust content of high-redshift
galaxies (Omont et al. 2001).  All of these scenarios would be
expected to produce highly inhomogeneous dust distributions, allowing
for identification with resolved high-sensitivity
thermal/submillimetre imaging.  A disturbed large-scale thermal
structure appears in some of our sources and one example is provided
by the Spitzer/MIPS 24\,$\mu$m image of HD~206773 shown in
Fig.~\ref{ima24.fig}. Colliding wind binaries would produce a dense
dusty region between the two stars, while the other two scenarios
would produce clumpy circumstellar shells. Finally, we wish to stress
the faintness of the dusty parsec-scaled halos, with a total optical
depth as low as $\simless 10^{-4}$ and dust masses below that of
Jupiter.

\subsection{Influence on ISM extinction or polarisation}

The Voshchinnikov sample of stars is frequently used to study dust in
the diffuse ISM (Hincelin et al. 2011, Boyarchuk et al. 2016, Zhukoska
et al. 2016 Shchekinov et al. 2017, Siebenmorgen et al. 2018). All of
the stars lie on sightlines with visual extinction of less than
2.5\,mag, colour excess of a few tenths of a magnitude
(Table~\ref{sample.tab}), and polarisation of a few percent, which is
assumed to be of interstellar origin. The observed polarisation curves
follow the Serkowski-law which is explained by dichroic absorption by
aligned dust grains and not by scattering on dust or free electrons
(Voshchinnikov 2012). We are therefore able to confirm that for our
subsample of 22 stars the polarisation originates in the ISM and not
in the faint dusty halo around the stars. Even for the far--IR excess
stars there is not sufficient circumstellar dust material available
that could significantly contribute: Their dust envelopes are too
optically thin to influence the observed reddening. The shallow
circumstellar dust halo also has no influence on the observed
polarisation, whether caused by absorption or by scattering.

\section{Conclusion}

By comparing models with high-sensitivity Spitzer/IRS spectra we have
found excess far-IR emission in 55\% of a sample of 22 massive stars
that are, with one exception, not known to be Be-stars and not connected
to clouds. The excess becomes apparent at wavelengths $\simgreat 10
\ \mu$m and is more pronounced in the far-IR. Black-body fits rule out
the possibility of background source contribution to the excess.  We
successfully applied wind and dust models to account for the detected
far-IR excess. Distinguishing or quantifying the contribution between
these scenarios requires high-sensitivity resolved imaging
observations at submillimetre wavelengths.

Spitzer/IRS offers the great advantage of observing mid-IR 
fine structure lines simultaneously with the far-IR continuum. This is
important because of the transient nature of the
excitation, as we demonstrate here. In stars with far-IR excess we detect Pf~$\alpha$ in 10
out of 12 sources while for  { non-}far-IR excess stars the detection rate
drops down to 50\% Pf~$\alpha$ and 33\% in Hu~$\alpha$, and where the
line intensities become an order of magnitude weaker.

By means of { echelle} high-resolution spectroscopy we detect (or
not) optical lines in H~$\alpha$, C~III and N~III
irrespective of the presence of a far-IR excess. However, the
detection rate of H~$\alpha$ is higher (60\%) for far-IR excess stars
than it is for{ non-}far-IR excess stars ($\sim 14\%$). We monitored
HD~149757 between 1993 and 2010 and find weak H~$\alpha$ emission in only one spectrum taken in
2006. For the Be-star HD~47839 we do not
detect H~$\alpha$ emission in a spectrum taken in 1996.  

For far-IR excess, which is due to dust emission, the bulk material is
located at parsec scales around the star in an extremely optically thin
($A_{\rm V, CS} < 500 \ \mu$mag) circumstellar halo. The circumstellar
dust halos are too faint to contribute to the observed extinction and
dichroic polarisation observed along the sightline towards these
sources. However, any significant dust population related to stars of
such high luminosity may require continuous replenishment to be
long-lived. Coronagraphic high-contrast imaging with SPHERE/IRDIS  for three of these stars rules
out the possibility that they  host debris-disk-like populations of
large rocky bodies whose collisions might provide a reservoir of
material.  Alternatively, the dust may be formed in-situ in the wind
or as a result of wind-wind interactions, possibly providing a channel
for dust formation at high-redshift. The existence of such dust may be detectable in a 24\,$\mu$m Spitzer image.


\begin{acknowledgements}
  We thank Robert Klement for discussions on Be-stars, T. D. Banas for
  supporting us in the SPHERE data reduction, and B. Altieri for help
  on the Herschel photometry. We are grateful to Nikolai Voshchinnikov
  for helpful comments and suggestions.  This research is based on
  data obtained from the ESO Science Archive Facility and in
  particular on observations collected under ESO programme
  095.C-0158(A).  This research has made use of the SIMBAD database,
  operated at CDS, Strasbourg, France. This work is based in part on
  observations made with the Spitzer Space Telescope, which is
  operated by the Jet Propulsion Laboratory, California Institute of
  Technology under a contract with NASA. This research is based on
  observations with AKARI, a JAXA project with the participation of
  ESA. This publication makes use of data products from the Two Micron
  All Sky Survey, which is a joint project of the University of
  Massachusetts and the Infrared Processing and Analysis
  Center/California Institute of Technology, funded by the National
  Aeronautics and Space Administration and the National Science
  Foundation. This publication makes use of data products from the
  Wide-field Infrared Survey Explorer, which is a joint project of the
  University of California, Los Angeles, and the Jet Propulsion
  Laboratory/California Institute of Technology, funded by the
  National Aeronautics and Space Administration.
\end{acknowledgements}


\end{document}